# White-light generating molecular materials: correlation between the amorphous/crystalline structure and nonlinear optical properties

Johannes Haust,[a] Jürgen Belz,[a] Marius Müller,[b] Benjamin Danilo Klee,[c] Jonathan Link Vasco,[c] Franziska Hüppe,[a] Irán Rojas Léon,[c] Jan Christmann,[c] Andreas Beyer,[a] Stefanie Dehnen,[c] Nils W. Rosemann,[d] Wolf-Christian Pilgrim,[c] Sangam Chatterjee,[b] and Kerstin Volz*[a]

[a]  J. Haust, Dr. J. Belz, F. Hüppe, Dr. A. Beyer and Prof. Dr. K. Volz
     Department of Physics and Materials Science Centre
     Philipps University Marburg
     Hans-Meerwein-Str. 6
     35032 Marburg, Germany
     E-mail: kerstin.volz@physik.uni-marburg.de
[b]  M. Müller, Prof. Dr. S. Chatterjee
     Institute of Experimental Physics I
     Justus Liebig University
     Heinrich-Buff-Ring 16
     35392 Gießen, Germany
[c]  Dr. B. D. Klee, J. Link Vasco, Dr. I. Rojas Léon, J. Christmann, Prof. Dr. S. Dehnen and Prof. Dr. W. C. Pilgrim
     Department of Chemistry and Materials Science Centre
     Philipps University Marburg
     Hans-Meerwein-Str. 4
     35032 Marburg, Germany
[d]  Dr. N. W. Rosemann
     Karlsruhe Institute of Technology
     Light Technology Institute
     Engesserstraße 13
     76131 Karlsruhe, Germany

**Abstract:** Amorphous materials are integral part of today´s technology, they commonly are performant and versatile in integration. Consequently, future applications increasingly aim to harvest the potential of the amorphous state. Establishing its structure-property relationship, however, is inherently challenging using diffraction-based techniques yet is extremely desirable for developing advanced functionalities. In this article, we introduce a set of transmission electron microscopy-based techniques to locally quantify the structure of a material. This unique approach allows to clearly identify the spatial distribution of amorphous and crystalline regions and to quantify atomic arrangements of amorphous regions of a representative model system. We study an ensemble of well-defined, functionalized adamantane-type cluster molecules exhibiting exceptionally promising nonlinear optical properties of unclear origin. The nanoscopic structure for three model compounds ([(PhSn)$_4$S$_6$], [(NpSn)$_4$S$_6$], [(CpSn)$_4$S$_6$]) correlates with their characteristic optical responses. These results highlight the advantageous properties of amorphous molecular materials when understanding the microscopic origin.

## Introduction

Amorphous materials are ubiquitous in today's technology. Applications range from amorphous silicon in photovoltaics to metal-based amorphous alloys for data storage or highly elastic metallic glasses.[1,2,3] Examples for recent developments include supercapacitors, electrolytic water splitting, and battery materials.[4,5,6] The amorphous materials show widely superior properties compared to their crystalline counterparts in their respective applications. The key for understanding and enhancing any desirable features, in general, is understanding the microscopic structure, or rather, the absence thereof in such materials.[7] Such an endeavor is particularly challenging as diffraction based techniques rely at least on pseudo-periodicity, and, commonly, lack sufficient spatial resolution indispensable in light of the continuous trend towards miniaturization. These challenges demand developing and applying advanced materials characterization procedures suitable for understanding aperiodic matter with sufficient spatial resolution while preserving the structure of the amorphous solid without causing extreme (radiation) damage during investigation.

A prototypical model system highlighting such necessities are ensembles of adamantane-type molecular clusters of the general formula [(RT)$_4$S$_6$] (R = organic substituent; T = Ge, Sn) which feature exceptional nonlinear optical properties of unclear origin. For selected R and T, these can exhibit directional white-light emission driven by a continuous-wave laser diode.[8] The immense application potential for such light sources is widespread. The initial observation consequently triggered studies of various adamantane-type clusters featuring different inorganic or organic core atoms and organic substituents with respect to their nonlinear optical properties.[8,9,10,11,12] The white-light generating (WLG) mechanism has tentatively been attributed to the presence of electron-rich organic substituents, their interaction with the driving electromagnetic field, and the absence of any long-range structural order in these materials.[13] The true microscopic nature of the process, however, remains undecided as well as any correlation with the arrangement of the clusters within the materials since their amorphous/crystalline structure on the relevant length scales remain unclear itself. Therefore, a key question for the understanding of the origin of the WLG mechanism is the quantitative understanding of the amorphous state of the materials with high spatial resolution. The latter should reveal any nanoscale local order and/or disorder including its influence on the nonlinear optical response. This clearly exceeds previous advanced structural investigations of any of these materials. ([(PhSn)$_4$S$_6$]), for example, has been characterized using synchrotron X - ray diffraction.[14,15] However, two major drawbacks of using X - rays for structure quantification are the low diffraction cross-section and the limited spatial resolution. Therefore, X - ray diffraction is generally unsuited for investigations of nanoscale volumes, where other methods like transmission electron microscopy (TEM) have proven to be of great use.[16] The electron-matter interaction is much stronger compared to X - ray diffraction and an electron probe can be used to irradiate areas ranging from a few nanometers to many micrometers.[17]

In this work, we exemplarily investigate different molecular cluster materials, [(PhSn)$_4$S$_6$], [(NpSn)$_4$S$_6$], and [(CpSn)$_4$S$_6$] (Ph = phenyl, Np = naphthyl, Cp = $\eta^1$ - cyclopentadienide), where the respective substituents have been varied systematically to explore the nature of the WLG process.[13] We find highly different nonlinear optical properties for the three materials, despite their nominally identical inorganic cores. Hence, these compounds are extremely well suited to showcase the application potential of electron diffraction - benchmarked by synchrotron X - ray diffraction for one selected compound - and microscopy experiments for structural characterization of amorphous materials and mixtures thereof with crystalline counterparts.

## Results and Discussion

Macroscopically, the three cluster compounds are white powders, where X-ray diffraction experiments identify their nanostructure as amorphous, i.e., no clear peaks are found in any of the diffraction patterns.[10,11] Their nonlinear optical properties, however, differ significantly, as will be described in more detail later. Most intriguingly, [(NpSn)$_4$S$_6$] shows second-harmonic generation rather than white-light generation, which is surprising considering the similarities to [(PhSn)$_4$S$_6$]. We will employ advanced electron microscopy, namely scanning precession electron diffraction (SPED), where we scan an ~1.5 nm sized precessing probe across the specimen, to reveal the distinct differences of the two compounds. TEM´s inherent local information circumnavigates any challenges imposed by small volumina of crystalline inclusions. These, however, pose a significant challenge for structure determination via X-ray diffraction, which lacks this high spatial resolution. SPED selectively addresses meaningful regions and collects sufficient separate datasets. Moreover, it distributes the total electron dose to a larger effective region with precise control also enabling the investigation of highly radiation-sensitive specimen as our molecular cluster compounds. Notably, the molecular structure of the adamantane-based clusters is ridged, i.e., structurally well-defined molecules assemble randomly and constitute the amorphous powder. This peculiarity might result in specific electron diffraction signatures. Therefore, we performed electron diffraction simulations to optimize the geometric parameters of the microscope beforehand (for details see SI 2, Figures S2, S3 and S4).

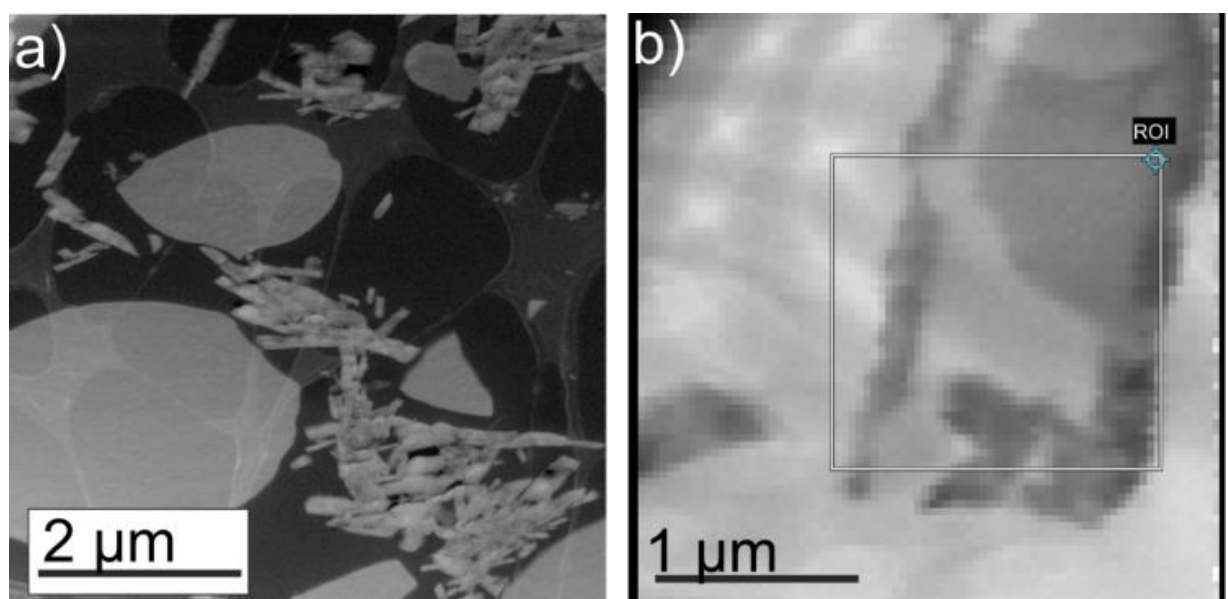

**Figure 1.** a) High angle annular dark field (HAADF) STEM overview images showing [(NpSn)$_4$S$_6$] particles of different size (bright regions) embedded in an epoxy matrix (black) on lacey carbon support (dark grey). b) Virtual bright field image generated from the SPED dataset showing an inverted contrast compared to the HAADF with the tin-sulphide cluster region being darker than the background. From this survey image a small sub set was generated for the indicated region of interest (ROI). Diffraction patterns from different regions of that data set are shown in Figure 2.

Figure 1a) shows the high angle annular dark field overview image for [(NpSn)$_4$S$_6$] acquired by STEM (scanning transmission electron microscopy). The tin-sulfide compound is displayed significantly brighter than the surrounding "transparent" epoxy matrix; the web-like structure of lacey carbon support is evident in the background. The compound itself is found in two different representations: large, µm-sized round particles and significantly smaller, rod-like particles. Figure 1b) shows the virtual bright field image reconstructed from the diffraction pattern intensities across the scanned area. The dashed box indicates the final SPED acquisition data set region that corresponds to the data shown in Figures 2 d-f.

Diffraction patterns (DP) are recorded and stored for all scan points in this region. Selecting specific pixels of the DPs for each scan point (Figure 2 a-c) generates virtual dark field (VDF) images proportional to their intensity (Figure 2 d-f). The majority of scan points (a, d) confirm the amorphous structure inferred from X-ray experiments. However, SPED clearly reveals nanoscale regions-of-order, i.e., exhibiting distinct diffraction spots (Figure 2 b-c, e-f).These crystallites (cyan and orange arrows), 50-150nm in size, are mostly found around the rod-like particles (cyan) as well as on the edges of the round particles (orange). The SPED analysis of other two compounds ([(PhSn)$_4$S$_6$] and [(CpSn)$_4$S$_6$]), however, does not hint any crystalline inclusions and confirms their amorphous state. Exemplary SPED data for [(PhSn)$_4$S$_6$], lacking any diffraction spots, is given in Figure S1 of the supporting information.

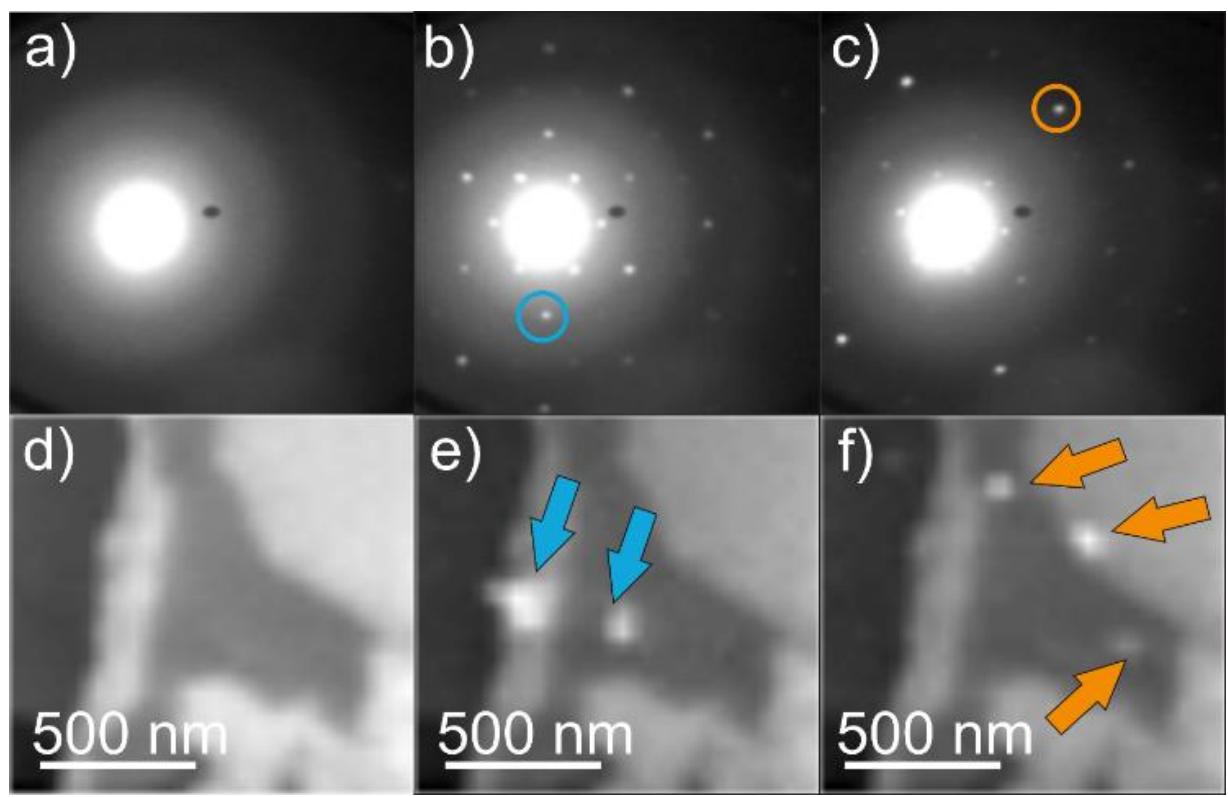

**Figure 2.** The diffraction pattern in a) originates from amorphous regions of the [(NpSn)$_4$S$_6$] material whereas b) and c) originate from regions highlighted in e) and f) respectively. By selecting specific regions in the diffraction pattern of the data set virtual dark field images can be generated. An arbitrary pixel taken from a) shows a dark field map of the scanned area, as depicted in d). In contrast, a region chosen on a diffraction spot like indicated in b) and c) highlights regions that generate these diffraction spots. It is apparent that the origin of the diffraction spots stems from small crystalline regions.

A detailed analysis of the electron diffraction data enables further characterization of the amorphous regions. Careful analysis of the transmission electron DPs (exemplarily shown in Figure 3 a)) is carried out by determining the structure factor S, which is a measure for the scattering power of a material. Details on the structure factor calculation and background subtraction routines are described in SI 3 of the supporting information. The comparison of the structure factor derived from experimental electron diffraction $S_{TEM}$ with the results from the X-ray diffraction data $S_{X-ray}$ is shown in Figure 3 b). The latter approach acts as benchmark for [(PhSn)$_4$S$_6$], confirming the validity and potential of the TEM-based approach. The diffuse diffraction rings of the TEM DP (Figure 3 a)) lead to peaks in the curve of $S_{TEM}$. This data underlines that the intramolecular order of the molecular clusters is preserved in the amorphous solid and that structural investigations of such a molecular cluster material with electron diffraction are indeed viable. As $S_{TEM}$ and $S_{X-ray}$ agree so well, it can also be concluded that even slight distortions of the inorganic core, which have been found by optimizing the structures with reverse Monte-Carlo (RMC) techniques for synchrotron X-ray diffraction[14,15], may also be detectable using TEM-based diffraction. It should be noted again that structure factor analysis is challenging, if not impossible, for X-ray diffraction, if crystalline inclusions are contained in the investigated volume. Hence, determination of bond lengths and accordingly quantification of the amorphous structure of molecular materials is possible using TEM, which will be elucidated for two further compounds in the following.

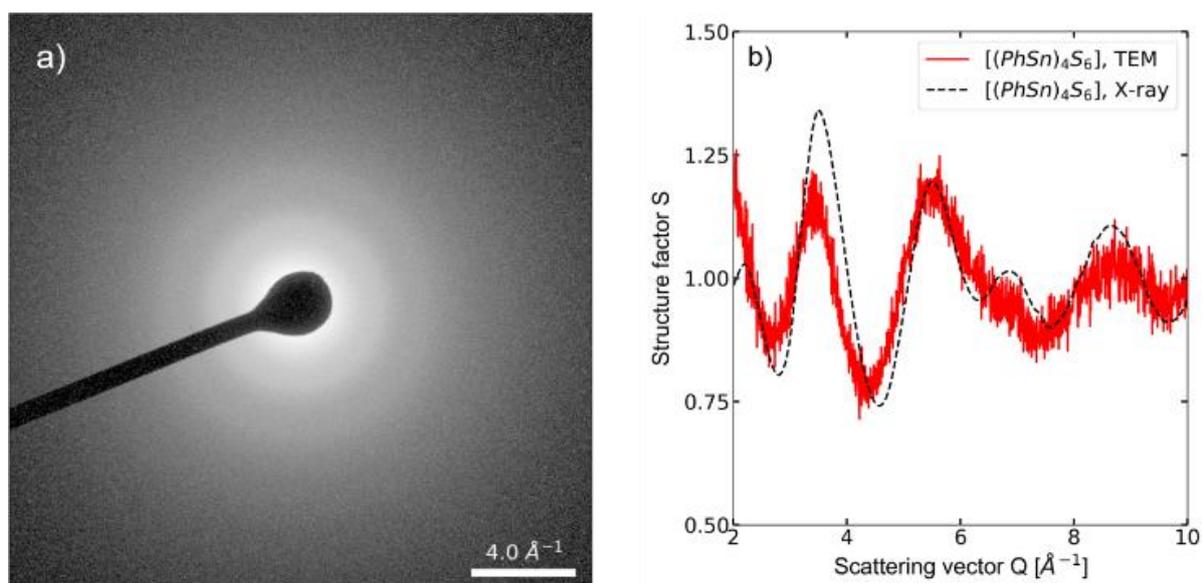

**Figure 3.** a) Raw experimental TEM diffraction pattern of a [(PhSn)$_4$S$_6$] specimen. In the diffraction pattern diffuse diffraction rings resulting from the amorphous structure of the sample are visible. b) Structure factor $S_{TEM}$ calculated from this diffraction image as described in the supporting information (please see SI 3.). The diffuse rings in the diffraction pattern are now visible as peaks in the structure factor. For comparison, the structure factor $S_{X-Ray}$ for [(PhSn)$_4$S$_6$] from X-ray diffraction (black dashed line, scaled with factor 0.25) is shown as well. The structure peaks of [(PhSn)$_4$S$_6$] in the X-ray data set are in excellent agreement with those from the TEM data, proving the suitability of TEM for the quantification of the amorphous structure of molecular materials.

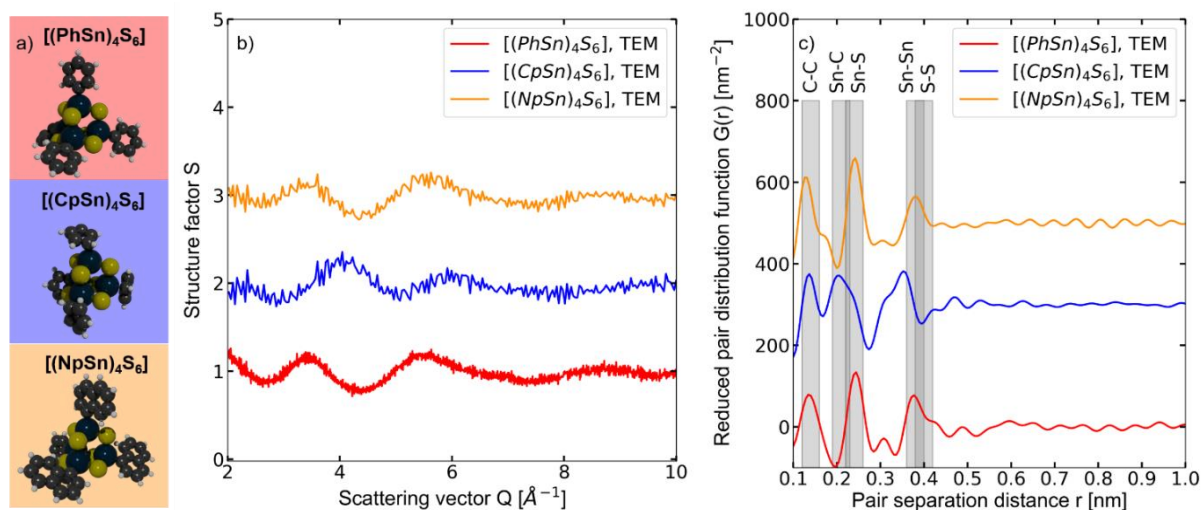

**Figure 4.** a) Visualization of the molecular compounds under investigation, based on the molecular structure as calculated using density functional theory.[11,13] b) Structure factors for [(PhSn)$_4$S$_6$], [(NpSn)$_4$S$_6$], and [(CpSn)$_4$S$_6$] (S for [(PhSn)$_4$S$_6$] was already shown in Figure 3 and is included here for better comparability) and c) Reduced pair distribution functions derived from the structure factors of the TEM data. The grey bars denote the nearest-neighbour distances of certain atoms. The [(NpSn)$_4$S$_6$] shows a symmetric pair distribution function, whereas for [(CpSn)$_4$S$_6$] and [(PhSn)$_4$S$_6$] the peaks of Sn - S, Sn - Sn and S - S are shifted and asymmetric pointing towards a slight distortion of the molecular building blocks. In regard with what has been shown, it should be noted that the diffraction pattern of the [(NpSn)$_4$S$_6$] has been taken from the amorphous region within the specimen.

The molecules under investigation are portrayed in in Figure 4 a). Their respective reduced pair distribution functions (PDF) G, giving average atomic distances in the solid, are calculated from the structure factor S, which is shown in Figure 4 b) for all compounds investigated (the calculation routines for S and G are explained in SI 3. of the supporting information). Clear differences in S between the compounds can be seen, which are also reflected in G, being the Fourier transform of S. By construction, G oscillates around 0 nm$^{-2}$. All reduced PDF curves are shown in Figure 4 c). They are offset vertically for better comparability. The PDF of [(PhSn)$_4$S$_6$] features prominent peaks at 0.15 nm, 0.24 nm and 0.38 – 0.4 nm. These are associated with bond lengths of C - C, S - Sn, Sn - Sn and S – S, respectively. The broad distribution of nearest neighbor distances also reflects the slight distortion of the inorganic core of the [(PhSn)$_4$S$_6$] molecules derived from X-ray experiments.[14,15] The reduced pair distribution function of [(NpSn)$_4$S$_6$] shows prominent peaks at similar positions, however, it is less asymmetric than the one of [(PhSn)$_4$S$_6$]. This indicates that the inorganic molecular core of [(NpSn)$_4$S$_6$] is less distorted (please note that the C - C bonding signal for this molecule is asymmetric as expected from the C-correlations of the Np group). In the case of the [(CpSn)$_4$S$_6$] molecule, the peaks corresponding to the atomic distances are shifted to slightly lower values and even more asymmetric than for the [(PhSn)$_4$S$_6$], hinting a distortion of the molecular core being present for this compound as well. These differences in atomic distances, thus, might be a reason for the highly different nonlinear optical responses or for the different solidification behavior of the compounds, respectively, as explained in the following

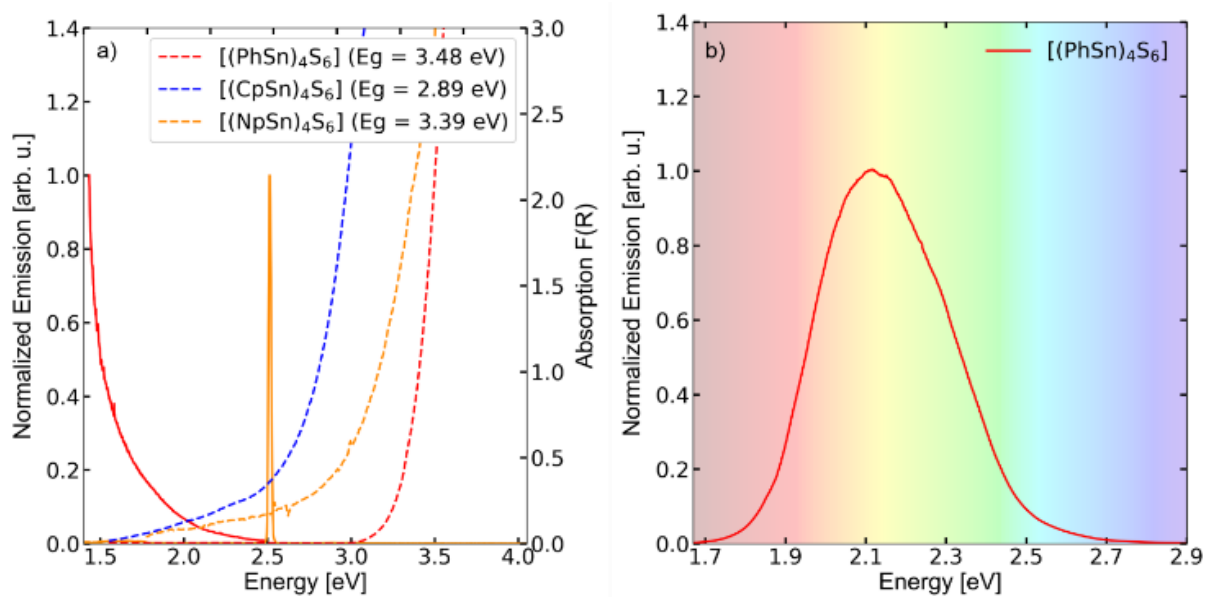

**Figure 5**. a) Nonlinear optical response (solid lines) following excitation at 1.265 eV and linear absorption spectra (dashed lines) for [(PhSn)$_4$S$_6$] (red), [(NpSn)$_4$S$_6$] (dark orange), and [(CpSn)$_4$S$_6$] (blue). The band gap energies are given in the legend, cf. Figure S5. The intense, nonlinear continuum response of [(PhSn)$_4$S$_6$] can be clearly seen ranging up to energies of 2.5 – 3 eV. [(NpSn)$_4$S$_6$] shows SHG around 2.5 eV, whereas [(CpSn)$_4$S$_6$] exhibits no nonlinear response when excited under similar conditions than the other molecular compounds. b) For better visualisation the nonlinear optical response of [(PhSn)4S6] is scaled with the photometric eye sensitivity.

The three molecular compounds under investigation show distinctly different optical and electronic properties, despite their molecular similarity. Intriguingly, the band-gap energies vary significantly: 3.48 eV, 3.39 eV, and 2.89 eV for [(PhSn)$_4$S$_6$], [(NpSn)$_4$S$_6$], and [(CpSn)$_4$S$_6$], respectively, see Figure 5. The values are derived according to Tauc's method adapted for condensed matter semiconductors assuming allowed, direct-gap transitions, see Figure S5.[18] [(PhSn)$_4$S$_6$] shows white-light emission observable by the bare eye for an excitation at 980 nm (1.265 eV).[8,9,13] The spectral data are shown scaled with the photometric eye sensitivity in Figure 5 b) for better visualization. This corroborated the previous attributions of WLG to purely amorphous nature of the compounds as prerequisite. Notably not only [(PhSn)$_4$S$_6$] but also [(CpSn)$_4$S$_6$] is a purely amorphous compound. Here, one has to consider that the latter compound has the lowest absorption edge energy of the materials under investigation. Tentatively, we find enhanced two-photon absorption into the pronounced tail states below the gap energy. This second-order nonlinearity associated with the imaginary part of the dielectric tensor is competitive with second-harmonic generation and can easily lead to breaking of bonds and, thus, transformation of the compound.

[(NpSn)$_4$S$_6$], however, exhibits a mixture of amorphous and crystalline regions, with the amorphous volume exceeding the crystalline volume by far. Notably, the amorphous regions in [(NpSn)$_4$S$_6$], contain molecular compounds with undistorted inorganic core which apparently is related to the likelihood for crystallization. The second-order optical nonlinearity observed for [(NpSn)$_4$S$_6$] can thus be explained by the vast amount of symmetry breaks at internal interfaces and surfaces in this polycrystalline phase – despite the inversion-symmetric structure of its crystalline domains - allowing second-harmonic generation.[19]

## Conclusion

In summary, we show that transmission electron microscopy is a very suitable technique for the quantitative structural investigation of amorphous materials, providing the indispensable extreme spatial resolution. To show its potential, we establish the structure-property relationships for a series of adamantane - type molecular clusters of the general formula [(RSn)$_4$S$_6$] (R = organic substituent) and benchmark the method against synchrotron-based X - ray analysis for the intramolecular bond lengths. The structural properties are in excellent agreement with characteristic behaviors, explaining the observed optical responses. Materials showing the intense, directed white-light emission need to be amorphous. Interfaces in the compounds, which are a result of crystalline/amorphous mixtures, inhibit the white-light generation process. These materials feature second harmonic generation instead, which is expected for materials lacking inversion symmetry. Interestingly, we observe slight distortions of the inorganic core for all materials, which are completely amorphous, pointing towards an influence of symmetry breaks on the observed properties. The wide availability of electron microscopes and demonstrated lateral resolution render this technique useful for extended screening studies as well as structure determination of non - crystalline condensed matter.

## Experimental Section

The synthesis of compounds comprising inorganic adamantane-type clusters of the general type [(RT)$_4$S$_6$] is commonly performed by reacting corresponding organotetreltrichlorides RTCl$_3$ with a sulfide source, usually alkali metal sulfides or (Me$_3$Si)$_2$S, in polar solvents (e.g., CH$_2$Cl$_2$).[20,21,22,23,24,25,26,27,28,29,30,31,32] Figure S6 illustrates the course of the reaction. Depending on the organic substituent, the products either precipitate as colorless, amorphous powders or form colorless crystals. For R = Ph, Np, Cp, the products are not single-crystalline. They have been synthesized according to the corresponding literature protocols.[11,13,28]

The amorphous powder material was embedded in an araldite epoxy for TEM preparation, cast in standard ultramicrotomy (UM) capsules and cured at 60 °C. In order to achieve high quality UM cuts, the specimen was trimmed to 200 – 300 µm with a trapezoid-like shape. The cutting was carried out in ambient conditions with a standard 45° wet cutting diamond knife. The resulting ribbons with a nominal thickness of 50 nm were transferred to clean lacey-carbon coated TEM grids. This procedure is further illustrated and explained in Figure S7.

TEM measurements were carried out using a conventional JEOL JEM - 3010 at 300 kV equipped with a TVIPS X416F-ES camera, which provides single-electron sensitivity. In order to avoid unnecessary contamination, inevitable adsorbents were polymerized by "beam showering" a neighboring region of the investigated one for about 30 min with high electron flux, effectively reducing the number of mobile contaminants. The microscope was operated at low-dose conditions, strongly assisted by the high sensitivity camera available. Hereto, we used the smallest condenser aperture available and the most parallel, wide spread beam. Dose rates for the measurements and locating images were in the order of $10^{-4}$ e/Å² for the low magnification mode and about $10^{-2}$ e/Å² for higher magnifications. The specimen regions were chosen such that there was neither epoxy nor lacey carbon support film within the selected diffracting area of about 0.03 µm². Measurements were carried out under room temperature conditions in order to provide comparable data to the previous X - ray diffraction studies.[14,15] For further reference the approach is illustrated in Figure S8 a - d.

Scanning precession electron diffraction (SPED) measurements were undertaken using the NanoMEGAS P2010 beam scanning/precession system installed at a double aberration-corrected JEOL JEM - 2200FS. This system creates a focused convergent probe with variable precession angle. In this work, we used an angle of about 1.0 degree. The probe is scanned across the specimen and the diffraction plane is recorded for every scan point by a camera pointing at the built-in phosphorous screen of the microscope. The pixel information therefore corresponds to the slightly convergent (about 0.8 mrad) diffraction pattern at a camera length of about 53 cm and originates from the probed area. The spatial resolution is determined by either the scan point resolution or the physical size of the probe which was measured to be about 1.8 nm for this experiment.

Electron diffraction simulations of [(PhSn)$_4$S$_6$] were carried out using the in-house developed *STEMsalabim* code to assess the electron diffraction of the molecular cluster materials and guide the experiments.[33] As input a cell was used, which had been derived by RMC modelling from previous synchrotron-based X-ray diffraction investigations.[14,15] This cell has a size of about 5.58 nm x 5.58 nm x 5.58 nm and contains 216 [(PhSn)$_4$S$_6$] molecules and therefore 11,664 atoms and their respective positions.

Diffuse reflectance spectroscopy in a tailored setup yielded the linear absorption according to the Kubelka Munk formulism.[34] White light from a xenon arc lamp or an incandescent source was monochromatized and focused into an integrating sphere using mirror optics. The scattered radiation was detected perpendicular to the excitation using lock-in technique; simultaneous measurements of the incident radiation compensates fluctuations.

For measurements of the nonlinear optical response the powder samples were dispersed on a borosilicate glass slide and placed inside a small vacuum cell pumped to pressures below $10^{-3}$ mbar. The continuous-wave 980 nm light from a laser diode was focused onto the samples using a 0.5 - NA reflective microscope objective in confocal geometry. The emission was imaged onto the entrance slit of a Czerny - Turner - type spectrometer where thermoelectrically cooled silicon charge-coupled device camera detects the spectrum; a heat absorbing filter suppressed the residual pump.

## Acknowledgements

Author 1 and Author 2 contributed equally to this work. We gratefully acknowledge support from the German Research Foundation (DFG) in the framework of the research unit FOR 2824. SC acknowledges the Heisenberg program under contract CH660/08.

**Keywords:** structure characterization amorphous/crystalline materials, (S)TEM, PED, reduced PDF, white light

**Supporting Information**

# White-light generating molecular materials: correlation between the amorphous/crystalline structure and nonlinear optical properties

*Johannes Haust, Jürgen Belz, Marius Müller, Benjamin Danilo Klee, Jonathan Link Vasco, Franziska Hüppe, Irán Rojas Léon, Jan Christmann, Andreas Beyer, Stefanie Dehnen, Nils W. Rosemann, Wolf-Christian Pilgrim, Sangam Chatterjee, and Kerstin Volz\**

## SI 1. SPED data set of [(PhSn)$_4$S$_6$]

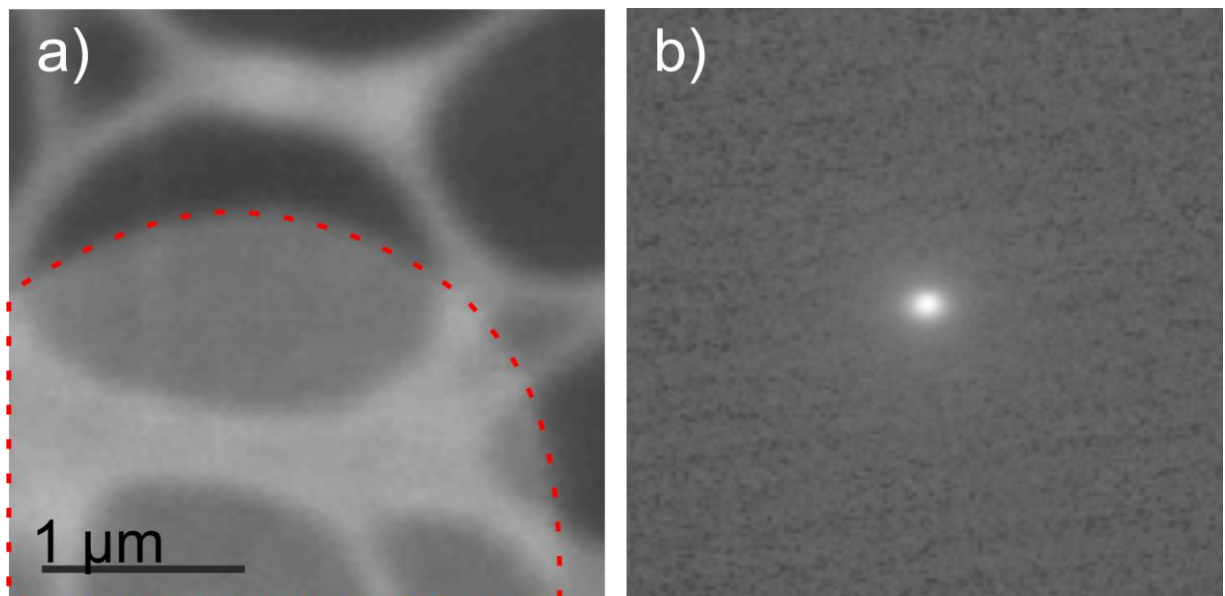

**Figure S1** A SPED virtual dark field of [(PhSn)$_4$S$_6$] is shown in a). The bright tin-based material is seen as a large round particle on a darker background (epoxy). For better visibility, the particle is encircled with a dashed, red line. In addition, the lacey carbon support can be seen as a web-like structure. b) A diffraction pattern of an arbitrary scan point is shown exemplarily to illustrate the absence of diffraction spots throughout the specimen.

In the main article, it is shown that the diffraction spots in [(NpSn)$_4$S$_6$] are from small crystallites of only a few tens to hundreds of nanometers in size. We verified the absence of small crystallites in [(PhSn)$_4$S$_6$] and [(CpSn)$_4$S$_6$] by SPED as illustrated in Figure S1. The virtual darkfield image generated by an arbitrary spot in the diffraction plane is shown in Figure S1 a). Therein, different domains are distinguishable. The web-like structure is the location where - in addition to the specimen sheet - the lacey carbon support film is located. The [(PhSn)$_4$S$_6$] particle is located in the lower part of the image and is generating a larger dark field signal (due to the higher atomic numbers) and is therefore brighter. In Figure S1 a) the respective region is encircled in red. The remaining dark regions are the embedding epoxy matrix. Exemplarily, a diffraction pattern from this dataset is shown in Figure S1 b), which does not contain any diffraction spots.

## SI 2. Electron diffraction simulations

An amorphous [(PhSn)$_4$S$_6$] cell, which is a result from the refinement of previous X - ray diffraction experiments, was used as input for the electron diffraction simulation. Using electron diffraction simulation the electron microscopic experimental parameters were optimized as well as those of the simulation.[1,2] As shown, their influence can be neglected.[3]

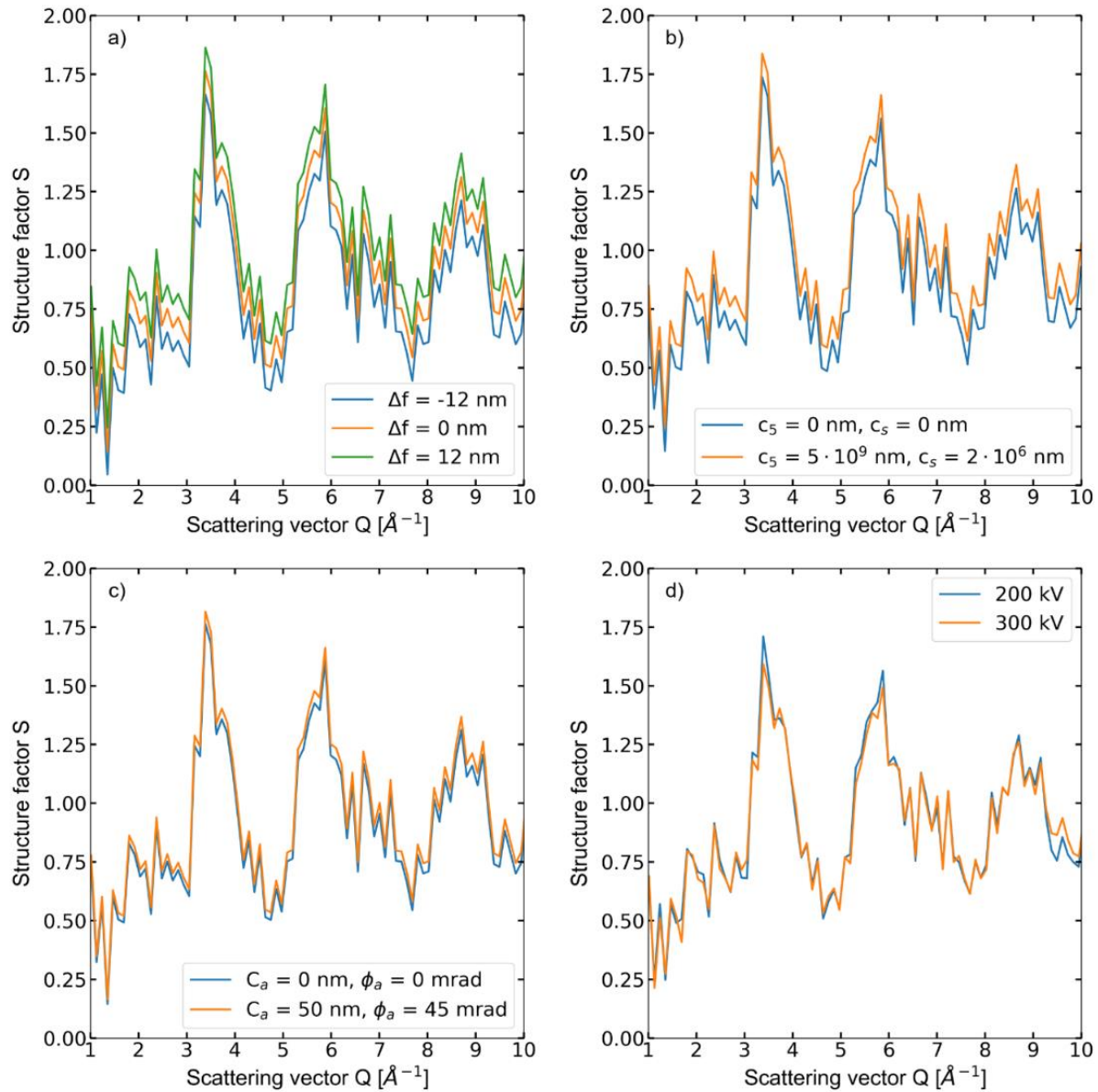

**Figure S2** Structure factors S calculated (as described in SI 3.) from electron diffraction simulations of an amorphous [(PhSn)$_4$S$_6$] cell with different geometric parameters of the microscope. a) Variation of the defocus Δf has no influence on S, the curves are shifted for visualization purpose only. b) Also the spherical aberration coefficients in third $C_S$ and fifth $C_5$ order have no influence on S, the curves are again shifted for visualization purpose only. c) The astigmatism coefficient $C_a$ and angle ϕ$_a$ have a small neglectable influence on S. d) Also the influence of the acceleration voltage is small and negligible.

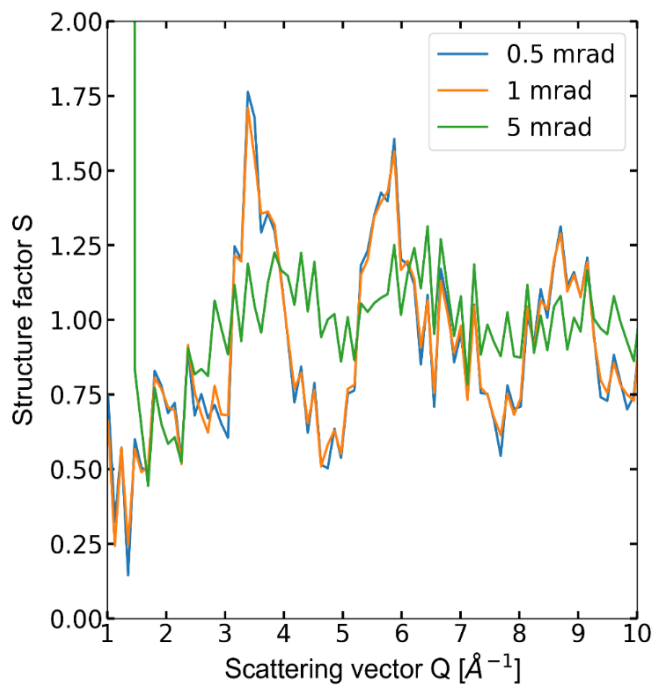

**Figure S3** Structure factors S from electron diffraction simulations of an amorphous [(PhSn)$_4$S$_6$] cell with different semi convergence angle α. The peaks in S broaden with larger α.

In Figure S3 the structure factors of simulations with different semi-convergence angles α of the probe of 0.5 mrad, 1 mrad and 5 mrad are shown. The main difference between the three simulations is the broadening of the peaks and the sudden increase of S the regime of small Q for α = 5 mrad. With larger α the diameter of the direct beam increases (in diffraction-space) and diffraction peaks with $Q < 2\pi \frac{\alpha}{\lambda}$ are therefore concealed through the direct beam.[4] Also, the peaks at larger Q, broaden due to the larger convergence angle.[4] This property effectively "blurs" structure in k-space, hence reducing the contrast. That restricts the tolerable convergence angle for structure determination.[5,6] In practice, one either has to use quasi parallel illumination or (de-)convolve the data recorded at higher convergence with the point spread function of the electron beam to investigate the structure of amorphous materials.[7]

In the following it is shown that the structure factor derived from electron diffraction simulation reproduces the X-ray structure factor, if an RMC input cell derived from X-ray diffraction data is used. In Figure S4 a) and b), the differential radial intensity and the derived structure factor for this diffraction simulation is shown. The peaks in both curves are resulting from the intramolecular structure of [(PhSn)$_4$S$_6$].

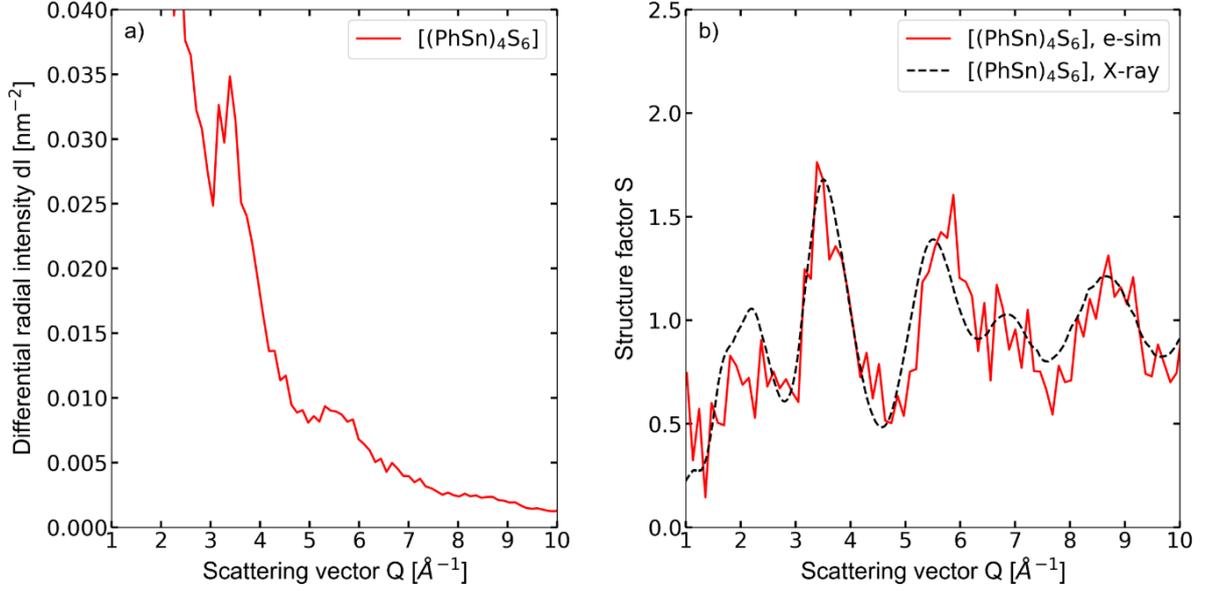

**Figure S4** a) Differential radial intensity (dI) from electron diffraction simulations of a [(PhSn)$_4$S$_6$] cell, which was refined from synchrotron X-ray diffraction data. b) Structure factor (S) calculated from the curve in a). dI as well S show peaks, which are a result of the intramolecular structure. The structure factor derived from synchrotron-based X - ray scattering (black curve, scaled by a factor of 0.5) is also shown. The distinct similarities between S from the electron diffraction simulation S$_{e-sim}$ and the one from the X - ray diffraction experiments S$_{X-ray}$ underline that structural investigation of the investigated class of molecular materials should indeed possible using electron diffraction also in experiment.

## SI 3. Structure Factor and Pair Distribution Function Calculation; Data processing and Background correction

The reduced pair distribution function (PDF) G(r) is a quantity, which is often used for describing the structure of amorphous materials.[6] It can be derived experimentally from diffraction data of X-rays, neutrons as well as electrons.[8] Hereto, the experimental diffraction data was processed in the following way:

The scattering vector Q was calculated from the electron scattering angle 2θ and the electron´s wavelength λ as given by [2]

$$Q = 4\pi \cdot \frac{\sin\left(\frac{2\theta}{2}\right)}{\lambda}. \tag{1}$$

From two-dimensional diffraction patterns, the radial sum of intensity I$_{sum}$ was calculated. In order to account for the changing momentum (Q)-space resolution, the so called differential intensity dI was obtained as [9]

$$dI(Q) = \frac{I_{sum}(Q)}{Q \cdot dQ}. \tag{2}$$

In order to calculate g(r), the mean scattering power has to be calculated. Because the investigated material is composed of several elements, the mean squared $\overline{f^2}(Q)$ is needed as well as the squared mean scattering factor $\overline{f}^2(Q)$ as explained in more detail elsewhere [6]

$$\overline{f^2}(Q) = \sum_i c_i \cdot f_i(Q)^2 \quad, \quad \overline{f}^2(Q) = \sum_i (c_i \cdot f_i(Q))^2, \tag{3}$$

where $c_i$ is the ratio of the respective element in the material. The atomic form factors $f_i$ were calculated according to the parametrization given in the literature.[2] The resulting structure factor S(Q) for kinematical diffraction is given by [6]

$$S(Q)=1+\frac{dI(Q)-\overline{f^2}}{\overline{f}^2}. \qquad (4)$$

The Fourier transformation of S is to then G(r) given as [6]

$$G(r)=\int_{Q_{min}}^{Q_{max}}(S(Q)-1)\cdot Q\cdot \sin(Q\cdot r)\,dQ. \qquad (5)$$

Peaks in G(r) give information about the distances between the particles.

The determination of S according to equation (4) is strictly valid for elastic single, i.e., kinematic, scattering. This applies per construction in the case of simulated data, because the energy loss of the electrons is not considered explicitly and the cell used is thin enough to safely neglect multiple scattering. However, both effects are present in the experiment. This implies that the mean scattering term is not well described by the mean scattering factor any more. One approach to overcome these difficulties is an empirical modeling of the background.[10] A positive portion of a Laurent-type series is used to model the combined effect of elastic, inelastic as well as multiple scattering. In this work, a similar approach is used, where this Laurent-type series

$$B_N(Q)=\sum_{i=2}^{N}\frac{c_i}{Q^i} \qquad (6)$$

together with a first scaling Term $\frac{c_0}{Q^{c_1}}$ is used, where $c_0, \ldots, c_N$ are fitting parameters. The final data correction is then performed as follows [10]:

$$dI_{corr}(Q)=\frac{\int_{Q_{min}}^{Q_{max}}\overline{f^2}(Q')dQ'}{\int_{Q_{min}}^{Q_{max}}\left(\frac{c_0}{Q'^{c_1}}dI_{raw}(Q')-B_N(Q')\right)dQ'}\cdot\left(\frac{c_0}{Q^{c_1}}dI_{raw}(Q)-B_N(Q)\right). \qquad (7)$$

Due to S(Q) → 1 when Q → ∞, $dI_{corr}(Q)$ should approach $\overline{f^2}(Q)$ in the regime of high Q values, which is here called "tail" of the respective quantity. The fit parameters are determined by minimizing the quadratic difference $\chi^2$ of the tails of $dI_{corr}(Q)$ and $\overline{f^2}(Q)$, respectively.[10]

$$\chi^2=\sum_{tail}\left(dI_{corr}(Q)-\overline{f^2}\right)^2. \qquad (8)$$

With this "corrected" intensity $dI_{corr}(Q)$ the structure factor S(Q) as well as the reduced pair distribution function G(r) can then be determined according to equation (4) and equation (5).

## SI 4. Optical spectroscopy

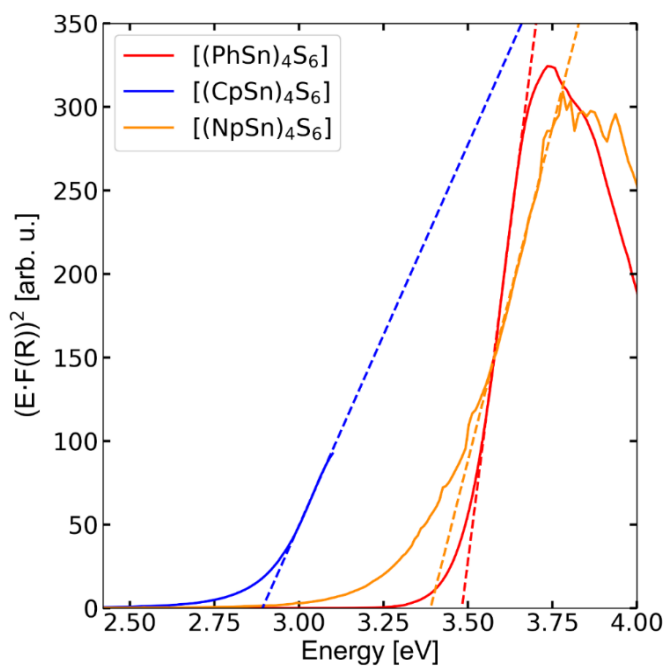

**Figure S5** Tauc plots for the three compounds [(PhSn)$_4$S$_6$], [(NpSn)$_4$S$_6$] and [(CpSn)$_4$S$_6$], assuming an allowed direct transition The linear fit curves are given as dashed lines.[11]

## SI 5. Chemical Synthesis

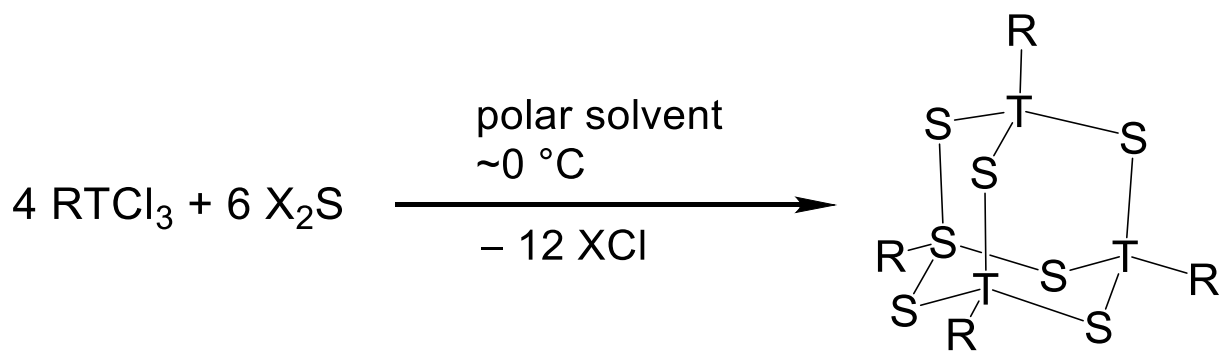

**Figure S6** General synthesis scheme for the preparation of cluster compounds of the type [(RT)$_4$S$_6$] (R = organic substituent; T = Ge, Sn; X = Na, K, SiMe$_3$).

## SI 6. TEM Specimen Preparation

The tin-sulfide clusters were produced as powders and stored under argon atmosphere until TEM specimen preparation. In order to achieve a reliable and constantly thin specimen thickness, we choose ultramicrotomy wet cutting for preparation of these specimens. Hereto the specimen is embedded in an epoxy that is compatible with high quality ultramicrotomy cutting and cured for several hours in dedicated capsules. The resulting epoxy stub is shown in Figure S7 a) from top illustrating the pyramidal shape of the curing capsule (not shown). The specimen (here: $[(PhSn)_4S_6]$) can be seen as a white powder in the transparent/yellowish epoxy matrix (Figure S7 a)). Following standard ultramicrotomy preparation guidelines the specimen region is trimmed to a trapezoid with a base length of about 300 µm (Figure S7 b)) in order to reduce the mechanical stress and provide a smooth surface with only small compression. The cuts were made with a sharp diamond knife aiming at a target thickness of 50 nm. The cuts typically attach to each other forming a ribbon floating on a water basin. This ribbon is transferred to a TEM compatible lacey carbon supported copper grid (Figure S7 c)).

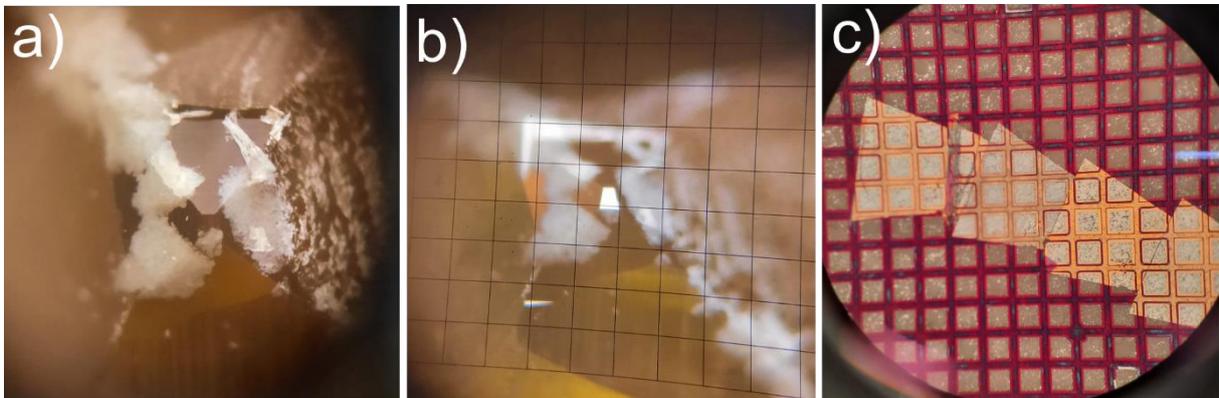

**Figure S7** a) The epoxy stub (yellowish) is pyramidally shaped and contains regions of the specimen powder (white). b) The base is trimmed to a trapezoidal shape with a base length of about 300µm. c) After cutting with a diamond knife these cuts float on a water basin and attach to each other forming a ribbon. This can be transferred to a TEM compatible lacey carbon coated grid

## SI 7. TEM Measurement Procedure

Figure S8 a) shows a low magnification TEM image with the round [(CpSn)$_4$S$_6$] particles embedded in epoxy on lacey carbon support film (overexposed for illustration only). Figure S8 b)-d) illustrates the measurement procedure. At first the region of interest is carefully approached with strong under focus for increased contrast at very low electron doses. Further on, a test exposure is made after stage stabilization (Figure S8 b)). A pre-centered selected area aperture and beam stop are inserted (Figure S8 c)) in order to restrict the diffraction to the preselected region and omitting exposing the direct beam to the camera. Finally, micrographs from the diffraction plane are recorded with higher exposures (0.2 e/ Å²) as shown in Figure S8 d) (excessive exposure for visualization only). In addition, as a final step the absence of drift is verified by taking another micrograph and comparison with the initial test exposure.

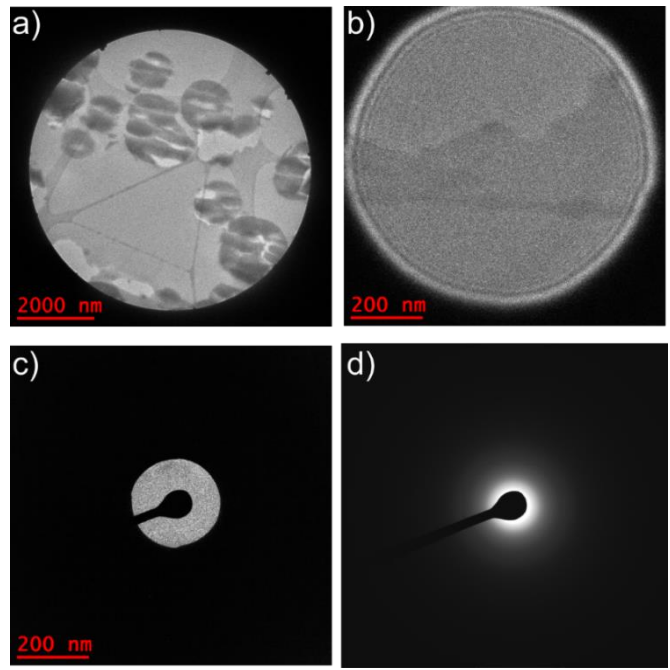

**Figure S8** a) The particles are located in low magnification and b) approached with low dose rates at higher magnification with a smaller illuminated region and strong under focus starting from pure epoxy regions. c) After identification of a suitable particle, a selected area aperture and a beam stop are inserted and the diffraction plane is recorded (d).